\documentclass[12pt]{article}
\usepackage{cmbright}
\usepackage[T1]{fontenc}
\usepackage[utf8]{inputenc}
\usepackage{amsfonts}
\usepackage[centertags]{amsmath}
\usepackage{amssymb,amscd,amsbsy}
\usepackage{array}
\usepackage{multirow}
\usepackage{graphicx}
\usepackage[colorlinks, linkcolor = black, citecolor = black, filecolor = blue, urlcolor = blue]{hyperref}
\usepackage{lineno}
\modulolinenumbers[5]
\usepackage{verbatim}
\usepackage{./upgreek}
\catcode`\@=11 
\def\parenbar{\mathpalette\p@renb@r}
\def\p@renb@r#1#2{\vbox{%
  \ifx#1\scriptscriptstyle \dimen@.7em\dimen@ii.2em\else
  \ifx#1\scriptstyle \dimen@.8em\dimen@ii.25em\else
  \dimen@1em\dimen@ii.4em\fi\fi \offinterlineskip
  \ialign{\hfill##\hfill\cr
    \vbox{\hrule width\dimen@ii}\cr
    \noalign{\vskip-.3ex}%
    \hbox to\dimen@{$\mathchar300\hfil\mathchar301$}\cr
    \noalign{\vskip-.3ex}%
    $#1#2$\cr}}}
\catcode`\@=12 
\def\nuan{\parenbar{\upnu}\kern-0.4ex}
\def\ean{\parenbar{\text{e}}\kern-0.4ex}
\def\muan{\parenbar{\upmu}\kern-0.4ex}
\def\tauan{\parenbar{\uptau}\kern-0.4ex}
\usepackage[square,numbers,sort&compress,comma]{natbib}
\usepackage[blocks,affil-it]{authblk}
\newcommand{\overbar}[1]{\mkern 1.5mu\overline{\mkern-1.5mu#1\mkern-1.5mu}\mkern 1.5mu}
\newcommand{\nue}{\upnu_{\text{e}}}
\newcommand{\numu}{\upnu_{\upmu}}
\newcommand{\nutau}{\upnu_{\uptau}}
\newcommand{\nuane}{\nuan_{\text{e}}}
\newcommand{\nuanmu}{\nuan_{\upmu}}
\newcommand{\nuantau}{\nuan_{\uptau}}
\newcommand{\anue}{\overbar{\upnu}_{\text{e}}}
\newcommand{\anumu}{\overbar{\upnu}_{\upmu}}

\newcommand\mysref[1]{Sec.\;\ref{#1}}
\newcommand\myfref[1]{Fig.\;\ref{#1}}

\newcommand\mytref[1]{Tab.\;\ref{#1}}
\newcommand{\diffunit}{\mathrm{GeV} \cdot \mathrm{cm^{-2} \cdot s^{-1}
    \cdot sr^{-1}}}
\title{An algorithm for the reconstruction of
  high-energy neutrino-induced particle showers and its application to
  the ANTARES neutrino telescope}
\author[1]{A.~Albert}
\author[2]{M.~Andr\'e}
\author[3]{M.~Anghinolfi}
\author[4]{G.~Anton}
\author[5]{M.~Ardid}
\author[6]{J.-J.~Aubert}
\author[7]{T.~Avgitas}
\author[7]{B.~Baret}
\author[8]{J.~Barrios-Mart\'{\i}}
\author[9]{S.~Basa}
% ...

\affil[1]{GRPHE - Universit\'e de Haute Alsace - Institut universitaire de technologie de Colmar, 34 rue du Grillenbreit BP 50568 - 68008 Colmar, France}
\affil[2]{Technical University of Catalonia, Laboratory of Applied Bioacoustics, Rambla Exposici\'o, 08800 Vilanova i la Geltr\'u, Barcelona, Spain}
\affil[3]{INFN - Sezione di Genova, Via Dodecaneso 33, 16146 Genova, Italy}
\affil[4]{Friedrich-Alexander-Universit\"at Erlangen-N\"urnberg, Erlangen Centre for Astroparticle Physics, Erwin-Rommel-Str. 1, 91058 Erlangen, Germany}
\affil[5]{Institut d'Investigaci\'o per a la Gesti\'o Integrada de les Zones Costaneres (IGIC) - Universitat Polit\`ecnica de Val\`encia. C/  Paranimf 1, 46730 Gandia, Spain}
\affil[6]{Aix Marseille Univ, CNRS/IN2P3, CPPM, Marseille, France}
\affil[7]{APC, Univ Paris Diderot, CNRS/IN2P3, CEA/Irfu, Obs de Paris, Sorbonne Paris Cit\'e, France}
\affil[8]{IFIC - Instituto de F\'isica Corpuscular (CSIC - Universitat de Val\`encia) c/ Catedr\'atico Jos\'e Beltr\'an, 2 E-46980 Paterna, Valencia, Spain}
\affil[9]{LAM - Laboratoire d'Astrophysique de Marseille, P\^ole de l'\'Etoile Site de Ch\^ateau-Gombert, rue Fr\'ed\'eric Joliot-Curie 38,  13388 Marseille Cedex 13, France}
% ...

\begin{document}
\maketitle

\begin{abstract}
  A novel algorithm to reconstruct neu\-tri\-no-in\-duced particle showers
  within the ANTARES neutrino telescope is presented. The method
  achieves a median angular resolution of $6^\circ$ for shower
  energies below 100\,TeV. Applying this algorithm to 6 years of data
  taken with the ANTARES detector, 8 events with reconstructed shower
  energies above 10\,TeV are observed. This is consistent with the
  expectation of about 5 events from atmospheric backgrounds, but also
  compatible with diffuse astrophysical flux measurements by the
  IceCube collaboration, from which 2 -- 4 additional events are
  expected. A $90\,\%$ C.L. upper limit on the diffuse astrophysical
  neutrino flux with a value per neutrino flavour of
  $\text{E}^2\cdot \upPhi^{90\%} = 4.9 \cdot 10^{-8}\,\diffunit$ is
  set, applicable to the energy range from 23\,TeV to 7.8\,PeV,
  assuming an unbroken $\text{E}^{-2}$ spectrum and neutrino flavour
  equipartition at Earth. 
\end{abstract}

\section{Introduction}
\label{sec:introduction}
With the discovery of a diffuse astrophysical neutrino flux by the
IceCube observatory located in the deep Antarctic ice, high-energy
neutrino astronomy has reported its first observation
\cite{Aartsen:2013bka,Aartsen:2013jdh,Aartsen:2014gkd}.  The
extraterrestrial origin of the flux has been established with high
significance~\cite{Aartsen:2014muf,icecube:HESE4,Aartsen:2016xlq}.
Although the sources of these high-energy neutrinos have not yet been
pinned down, it is expected that their identification will help to
elucidate the sites and mechanisms of baryonic acceleration, and will
play a key role in the discovery of the sources of Galactic and
extragalactic cosmic rays.

In neutrino telescopes in ice or water, a charged-current (CC)
interaction of a $\numu$ or $\anumu$ (in the following abbreviated to
$\nuanmu$) inside or around the instrumented volume creates a
relativistic muon whose long trajectory can, depending on its energy,
cross the entire detector and be detected by photomultipliers (PMTs)
through the induced Cherenkov light emission.  The event signature due
to neu\-tral-current (NC), and $\nuane$ and $\nuantau$ CC interactions
inside or close to the instrumented volume is however a particle
shower\footnote{with the exception of about 17\% of tau leptons
  decaying to muons that appear as track-like
  events~\cite{bib:PDG2016}.} 
(also often referred to as a shower-like or cascade event) 
with a characteristic longitudinal extension of a few meters that
increases logarithmically with energy. The particle shower constitutes
a Cherenkov light source which appears localised compared to the typical
distances between photosensors in neutrino telescopes.  This light
emission characteristic offers the opportunity to estimate the energy
released in a neu\-tri\-no-in\-duced shower more reliably than that of
muons, while the direction determination is more difficult and
generally results in a worse angular resolution.

A high-energy astrophysical neutrino flux has been observed and
characterised in several different analyses by IceCube. The
high-energy starting event analysis identifies
neu\-tri\-no-in\-teraction vertices of all flavours contained in the
detector volume. In 4 years of data taking, it has observed 54 events
from the entire sky, of which 39 have been identified as shower-like
with a typical directional resolution of about $15^{\circ}$
\cite{icecube:HESE4}.  A best-fit spectral index of
$\upGamma=2.58\pm 0.25$ is obtained, assuming a power-law flux model
$\text{dN}_{\upnu}/\text{dE}_{\upnu} = \upPhi_0\text{E}^{-\upGamma}$.
The flux normalisation at 100\,TeV of
$\upPhi_0 = 2.2\times 10^{-8}\,\diffunit $ is valid per neutrino
flavour, and for neutrinos yielding a deposited energy between
$60\,\text{TeV}$ and $3\,\text{PeV}$.  Recently, a complementary
measurement of an astrophysical neutrino flux has been achieved using
only CC muon neutrino events from the Northern sky. Using 6 years of
data, an astrophysical flux with a hard spectral index of
$\upGamma=2.13\pm 0.13$ and a normalisation at 100\,TeV of
$\upPhi_0 = 0.9\times 10^{-8}\,\diffunit $ has been found for neutrino
energies above roughly 200\,TeV \cite{Aartsen:2016xlq}. This result
shows a $3.3\,\upsigma$ tension with the normalisation value and soft
spectral index obtained in a fit combining different previous IceCube
analyses with mainly lower energy thresholds \cite{Aartsen:2015knd},
which could be indicative of a spectral break \cite{Aartsen:2016xlq}.
The measurements indicate that a substantial fraction of the flux must
be of extragalactic origin, while a Galactic contribution could be the
reason for the observed tension.  Exploiting the limited statistics of
the available astrophysical neutrino sample, first indications have
been put forward that the observed flux is anisotropic, being slightly
stronger and exhibiting a softer spectrum in the region of the Galaxy
in the Southern sky \cite{Neronov:2015osa,Neronov:2016bnp}.  The
$\nue:\numu:\nutau$ ratio is compatible with 1:1:1
\cite{Aartsen:2015knd}, consistent with expectations from charged
meson decays in cosmic-ray accelerators and 3-flavour neutrino mixing.
Dedicated searches for small-scale anisotropies in neutrino arrival
directions and for spatial correlations with known astrophysical
sources have not revealed statistically significant deviations from
the isotropy hypothesis
\cite{Aartsen:2016oji,Adrian-Martinez:2015ver,Adrian-Martinez:2014wzf,Aartsen:2014ivk}.

Given the tensions and uncertainties in the observations by IceCube,
it is important to provide additional measurements and complementary
sky coverage in the track-like muon neutrino and in the shower-like
all-flavour event channels. ANTARES is a neutrino telescope located in
the Northern Hemisphere which, despite having a significantly small\-er
volume than IceCube, has a comparable muon neutrino effective area at
TeV energies for observations of the Southern sky
\cite{Adrian-Martinez:2015ver}.  
ANTARES data have been used to set constraints on, e.g., the all-sky
diffuse muon neutrino flux~\cite{Aguilar:2010ab,Schnabel:2015ena}, the
strength of a possible Galactic component of the flux discovered by
IceCube~\cite{2014PhRvD..90j3004S}, and the possible neutrino flux
from the region of the Galactic Ridge
\cite{Adrian-Martinez:2016fei}. Furthermore, several searches for
clustering and large-scale anisotropies in the neutrino arrival
directions, as well as for temporal and/or spatial correlations with
known astrophysical sources have been carried out
\cite{antares_pointsources2,Adrian-Martinez:2013dsk,antares_fermibubbles,Adrian-Martinez:2014hmp}.

This paper presents a reconstruction algorithm for neu\-tri\-no-in\-duced 
particle shower events and reports on the first application of such an
algorithm to ANTARES data.  The reconstruction method has been
employed to search for a diffuse astrophysical neutrino flux using 6
years of data collected from 2007 to 2012. The ANTARES detector is
described in \mysref{sec:antares}.  The detector simulation and the
developed algorithm are presented in \mysref{sec:simulation} and
\mysref{sec:reconstruction}, respectively.  The data selection is
discussed in \mysref{sec:dataselection}, while the analysis method and
the discussion of systematic uncertainties can be found in
\mysref{sec:analysismethod}.

The results of the search are reported in \mysref{sec:results}, while
\mysref{sec:conclusion} summarizes and concludes the paper.  The
presented work is used as input to more advanced reconstruction
algorithms based on updated simulations which are in
development~\cite{Michael}.

\section{The ANTARES neutrino telescope}
\label{sec:antares}
The ANTARES neutrino telescope~\cite{antares_overview} is located in
the Me\-di\-ter\-ra\-nean Sea about 40\,km offshore from Toulon in a depth of
about 2500\,m, and comprises a three-dimensional array of 885 PMTs
housed inside glass spheres, denoted as optical modules (OMs)
\cite{antares_opticalmodule}. The OMs are attached to 12 readout
cables (lines), each holding 75 of these arranged in groups of three
on 25 storeys\footnote{The 12th line holds only 20 storeys with
  OMs. The remaining storeys house a test system for acoustic neutrino
  detection~\cite{Aguilar:2010ac}.}.  The vertical spacing between
storeys is 14.5\,m, while the horizontal spacing between lines
deployed in an approximately octagonal configuration is about 60\,m on
average.  The detector instruments a water mass of roughly
$20\,\text{Mt}$, but can be sensitive to neutrino interaction events
outside of this volume, depending on the distance of the neutrino
interaction point (vertex) to this volume, the neutrino direction and
the event light yield.  ANTARES is mainly sensitive to neutrinos of
TeV to PeV energies, with a threshold for astrophysical studies of
roughly $100\,\mathrm{GeV}$.

If the analogue output signal of a PMT reaches an amplitude
corresponding to a charge above a tunable threshold of typically
0.3\,photoelectrons (pe), the signal time and charge are digitised,
and this pair of values is denoted as a ``hit''~\cite{Aguilar:2010bw}.
Events are selected by different triggering algorithms
\cite{antares_daq} that causally connect hits in time and space.  The
achieved resolutions on the arrival time of photons at the PMTs,
measured with nanosecond precision~\cite{antares_timecalibration}, and
on the position and orientation of the OMs~\cite{antares_positioning},
as well as the low photon scattering probability in
seawater~\cite{Aguilar:2004nw}, allow for the reconstruction of the
triggered events with excellent angular resolution for muon neutrino
CC events \cite{AdrianMartinez:2011uh}.
 
Two different types of backgrounds have to be taken into account in
the event reconstruction algorithms and in the search for high-energy
astrophysical neutrinos.  The time variable photon emission by
deep-sea bioluminescent organisms and Cherenkov photons induced by
electrons from beta decays of radioactive potassium (${}^{40}$K) add
PMT hits unrelated to those caused by the detection of Cherenkov
photons from the passage of relativistic particles.  The second type
of background consists of events that are induced by atmospheric
neutrinos and muons produced in interactions of cosmic rays with the
Earth's atmosphere.  Using the Earth as a shield against the
atmospheric muon background, upward-going neutrinos are observed that
predominantly originate from the Southern sky due to the geographical
location of the telescope.

Individual upward-going atmospheric neutrinos are indistinguishable
from neutrinos of astrophysical origin, unless observed in temporal
and/or spatial coincidence with other cosmic messengers
\cite{Bonis:2016qol,Ageron:2011pe}.

\section{Simulation of signal and background}
\label{sec:simulation}
For the development of the shower reconstruction algorithm
and for the optimisation of the diffuse neutrino flux search, detailed
Monte-Carlo (MC) simulations of the detector response to both signal and
background events are used~\cite{Brunner:2003ke,Margiotta:2013jma}.

Some of the deep-sea environmental conditions typically change on a
time\-scale of a few hours. In particular, the optical background
rates, which are measured for each OM individually, can show
significant variations with time, and are of relevance for the data
acquisition and the detector efficiency.  In order to take these
variations into account, each data-taking period of a few hours
(denoted as a {\it run}) is simulated individually
\cite{Fusco:2016jvz}. The background is generated according to the
measured rates on each active OM, which are determined with a sampling
frequency of roughly 10\,Hz.  Additionally, PMT individual charge
calibrations and effective thresholds are used, and the simulated hit
time and charge is smeared.  Finally, the simulated events are
processed with the same trigger algorithms active during data
acquisition.
 
The generation of $\nuanmu$ and $\nuane$ neutrino interactions is
performed using the LEPTO~\cite{ingelman_lepto} package for deep
inelastic scattering processes and RSQ \cite{barr_phd} for resonant
and quasi-elastic processes using the CTEQ6-DIS~\cite{pumplin_cteq6d}
parton distribution functions.  The hadronisation is performed using
PY\-THIA / JET\-SET \cite{sjostrand_jetset}.  Interactions of $\nuantau$ are
not simulated and their contribution is estimated differently, as
discussed in \mysref{sec:analysismethod}.  In order to obtain
sufficient statistics at high energies, the $\nuane$ and $\nuanmu$
events are generated with a hard $\mathrm{E}^{-1.4}$ spectrum. A
reweighting procedure is employed to simulate different astrophysical
and atmospheric neutrino flux models from the generated events.

The generation of atmospheric muon events uses the MUPAGE
\cite{antares_mupage,antares_mupageupdate} package.  The propagation
of muons in water is achieved with MUSIC~\cite{antonioli_music}.  For
muon events, no reweighting procedure is used, but an integrated flux
corresponding to one third of the data-taking livetime is generated.

For hadronic showers induced by neutrinos with an energy below
$100\,\mathrm{TeV}$, each particle generated in the interaction and
its corresponding light emission is simulated with GEANT 3.21
\cite{cern_geant}.  Electromagnetic showers and their photon emission
are generated using parametrisations and precomputed probability
tables.  For neutrino events with energies above $100\,\mathrm{TeV}$,
hadronic showers are simulated using a one-particle approach, i.e. all
hadrons are replaced with an equivalent electron whose energy is
determined from that of the hadrons by an appropriate weighting
scheme.
 
In order to keep the computational cost of the simulation manageable,
two additional simplifications are introduced.  For photons generated
in particle showers, scattering processes are not taken into account,
and for $\nuanmu$ CC events with
$\mathrm{E}_{\upnu} > 100\,\text{TeV}$, Che\-ren\-kov photon emission from
the hadronic vertex shower is not simulated.  Both simplifications are
taken into account in the analysis by corrections and corresponding
systematic uncertainties, which are derived from dedicated simulations
and discussed in \mysref{sec:analysismethod}.

\section{Shower event reconstruction}
\label{sec:reconstruction}
For the selection and reconstruction of triggered events that contain
a shower, a dedicated maximum-likelihood-based reconstruction
algorithm has been developed.  It allows for the estimation of the
shower energy, of the interaction point and time, and of the direction
of the incoming neutrino.

In a pre-fit step, the shower position and time are roughly estimated.
To this end, hits caused mainly by unscattered light are selected by
considering only the earliest hit on each OM.  A $\chi^2$-fit scanning
for the time and position of the show\-er is done assuming a spherical
light source, and using only OMs on storeys with at least two hits
within $20\,\mathrm{ns}$.  As optical background processes, such as
${}^{40}$K decays or bioluminescence, induce mainly single
photoelectron hits, restricting the hit selection to coincidences with
a charge exceeding $1.2\,\text{pe}$ per hit ensures that this pre-fit
is performed on a sample dominated by signal hits.  This signal hit
selection has been developed with dedicated simulations including
scattering for photons induced by shower particles, and has been
verified by comparing measured and simulated hit time distributions.

In the next step, a new hit selection takes into account all hits in
the event again.  Hits are selected if their distance to at least one
storey with coincident hits or to the shower position estimated in the
previous step is lower than $50\,\mathrm{m}$. Additionally, the hit
time must be in a range of $\pm80\,\mathrm{ns}$ with respect to the
arrival time expectation assuming isotropic photon emission at the
estimated shower position.  The chosen value of the distance criterion
corresponds roughly to the seawater absorption length
\cite{YepesRamirez:2011zza} and prevents far-away background hits that
coincidentally fit to the isotropic light emission hypothesis from
being falsely selected.  If this procedure finds fewer than 5 hits in
total or hits on less than 3 lines, the event is discarded. The
remaining contamination from noise-induced hits has been estimated to
be about 1\,\%.

Refining the results of the pre-fit and based on this second hit
selection, the parameters of the shower are determined with two
consecutive maximum-likelihood fits.  Both fits make use of
precomputed probability tables that have been obtained using the
detailed MC simulations described in \mysref{sec:simulation}.  The
first maximum-likelihood fit determines the position and time of the
shower. It varies these shower parameters and evaluates the
precomputed probability for each selected hit, given its time and
position, to be due to Che\-ren\-kov photons emitted at the assumed
shower time and position.  The second fit determines the direction of
the incoming neutrino and the energy of the particle shower resulting
from the neutrino interaction, while fixing the start time and
position of the shower to the values found by the first fit. This
factorisation of the fitting procedure is possible due to the large
scattering length of seawater and due to the homogeneity of the
medium, which allows for the position reconstruction of the maximum
shower light yield independent of the shower
direction\footnote{cf. Ref.~\cite{Adrian-Martinez:2016fdl}, in
  particular Sec.~4.4.2.}.  This second fit is based on precomputed
and tabulated probabilities for hits to be due to Cherenkov photons
emitted in a particle shower with given energy, time and position, and
induced by a neutrino with given direction.  The three-dimensional
probability table depends on the photon emission angle, the total
photon yield emitted by the shower, and the energy of the shower. The
photon emission angle is defined as the angle between the direction of
the incoming neutrino and a straight line from the shower position to
the hit OM. 
The shower charge $\text{c}_{\text{shower}} $, in units of
photoelectrons and with typical values of about $10^8$\,pe for 10\,TeV
shower energy, is used as a proxy for the total light yield from the
shower and is defined as
$\text{c}_{\text{shower}} = \text{c}_{\text{hit}} \cdot
\text{e}^{\frac{\text{d}}{\uplambda_{\text{w}}}} \cdot
\frac{1}{\upalpha} \cdot \frac{4\uppi
  \text{d}^2}{\text{A}_{\text{OM}}}$,
where $\text{c}_{\text{hit}}$ is the measured charge of the hit with a
maximum value of about 25\,pe, $\uplambda_{\text{w}}$ is the
attenuation length of seawater \cite{Aguilar:2004nw} and $\upalpha$ is
the incidence-angle-dependent photon-detection probability of an OM.
The last factor relates the OM cross-section
$\text{A}_{\text{OM}}$~\cite{antares_opticalmodule} to the total
surface of a sphere defined by the radial distance d of the shower
position to the OM.  The definition of the parameter
$\text{c}_{\text{shower}}$ was chosen to make the shower energy
estimate approximately independent of the detected light yield,
allowing for the reconstruction of events in which the emitted light
partly escapes the sensitive volume of the detector.

In the search for astrophysical neutrinos described later, a quality
cut on the likelihood of the vertex fit ({\it vertex-quality cut}) is applied.  It
aims at optimising the signal to background ratio by efficiently
selecting neutrino-induced show\-er events while vetoing atmospheric
muons.  Applying this cut yields a 3 (6)\,$\mathrm{m}$ median position
resolution for the neutrino interaction vertex for events with a
MC shower energy of $100\,\mathrm{GeV}$
$(1\,\mathrm{PeV})$. In particular, for high shower energies, this
resolution is dominated by the distance between the interaction vertex
and the position of the shower light yield maximum. The MC
shower energy is defined by the fraction of the neutrino energy
deposited at the vertex, thus contributing to the shower light yield.
For $\nuane$ CC events, it is equivalent to the neutrino
energy, while it is lower by the energy of the escaping neutrino for
NC events.

\begin{figure*}[]
\centering
 \includegraphics[width=0.49\textwidth]{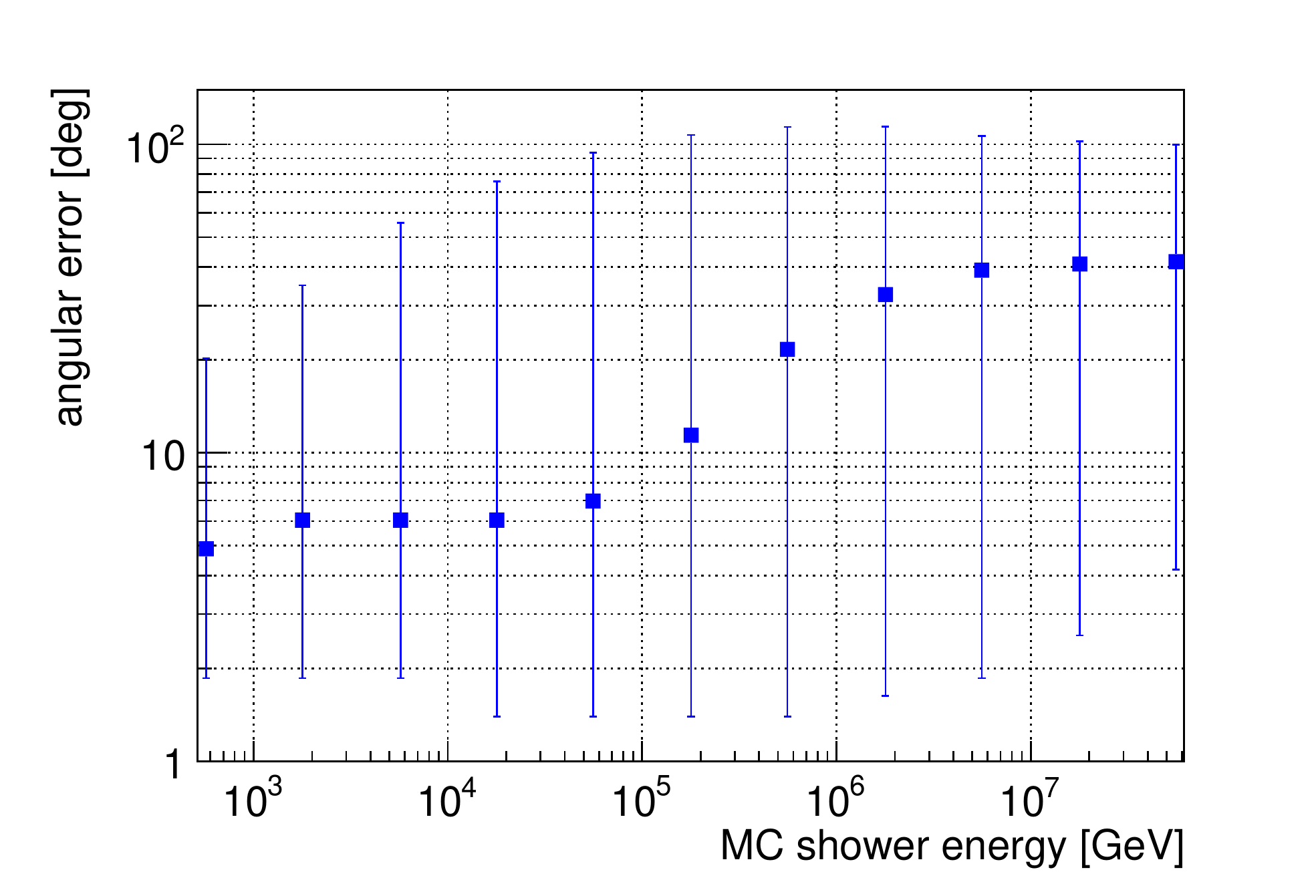}
 \includegraphics[width=0.49\textwidth]{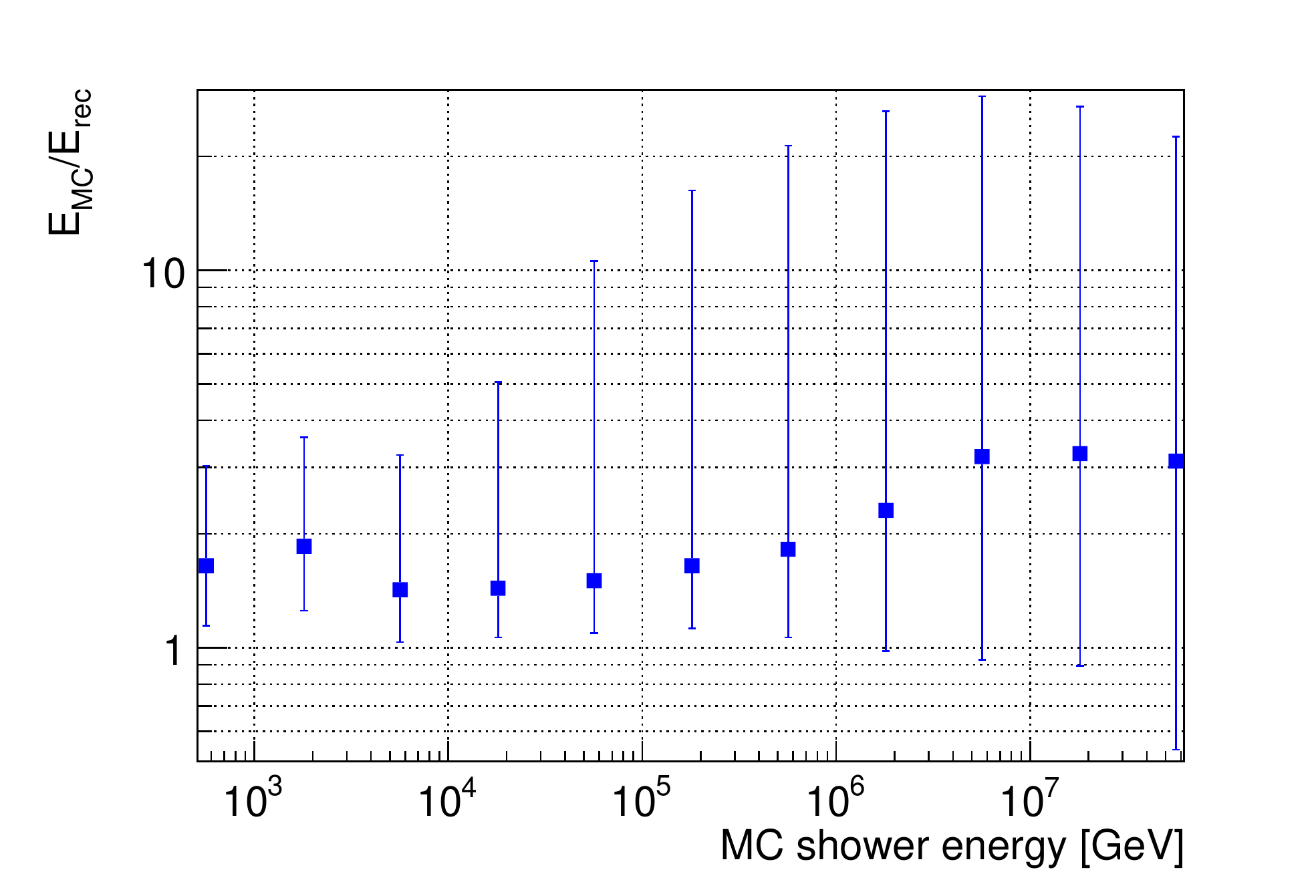}
 \caption{Left: Angular error of the direction reconstruction for
   shower-like neutrino events as a function of the MC shower
   energy. Right: The ratio of the MC and the reconstructed shower
   energy, as a function of the MC shower energy. Blue squares denote
   the median of the distributions. The lower and upper end of the
   vertical bars in both figures show the 10\,\% and 90\,\% quantiles
   of the distributions, respectively.}
\label{fig:reco-reso}
\end{figure*}
The distribution of the angular error on the neutrino direction in
\myfref{fig:reco-reso} (left) shows a median value of about $6^\circ$
for shower energies up to $100\,\mathrm{TeV}$, and worsens to about
$25^\circ$ ($40^\circ$) at $1\,\mathrm{PeV}$
($10\,\mathrm{PeV}$). This is a consequence of the stronger light
yield at higher energies that saturates the detector and increasingly
impedes the efficient recognition of the emission direction of
Cherenkov light from the shower particles.

The ratio between the MC and the reconstructed energy
$\mathrm{E_{MC}}/\mathrm{E_{rec}}$, characterised by its median value
as well as the 10\% and 90\% quantiles, is depicted as a function of
shower energy in \myfref{fig:reco-reso} (right).  The median value
stays below 2 for shower energies up to $1\,\mathrm{PeV}$, and
increases to about 3 at $10\,\mathrm{PeV}$.  While 90\% of the events
are reconstructed with a ratio $\mathrm{E_{MC}}/\mathrm{E_{rec}}$ up
to 4 for energies below 10\,TeV, the distribution widens significantly
up to PeV energies, again as a consequence of the light yield
saturating the detector.

\begin{figure*}[]
\centering
 \includegraphics[width=0.49\textwidth]{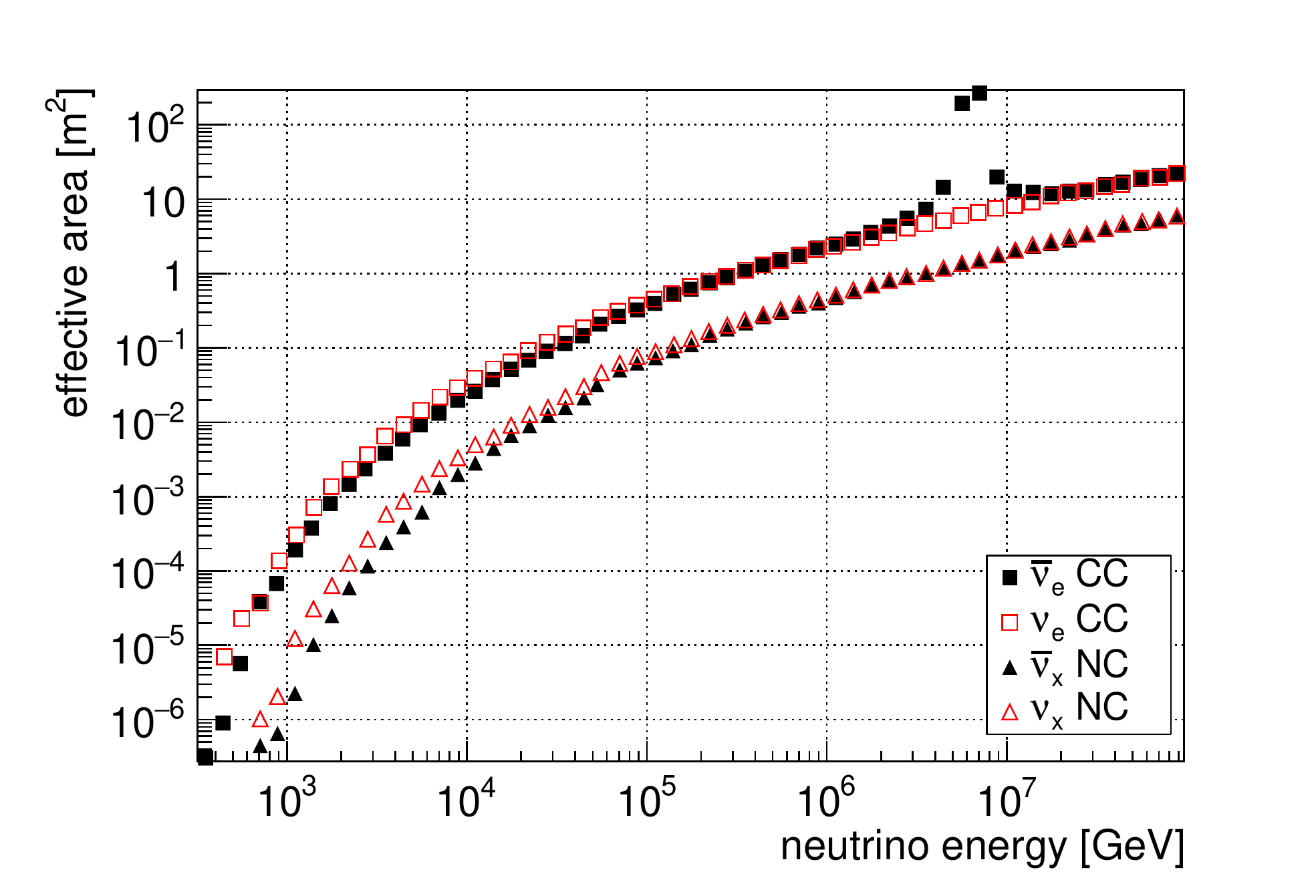}
 \includegraphics[width=0.49\textwidth]{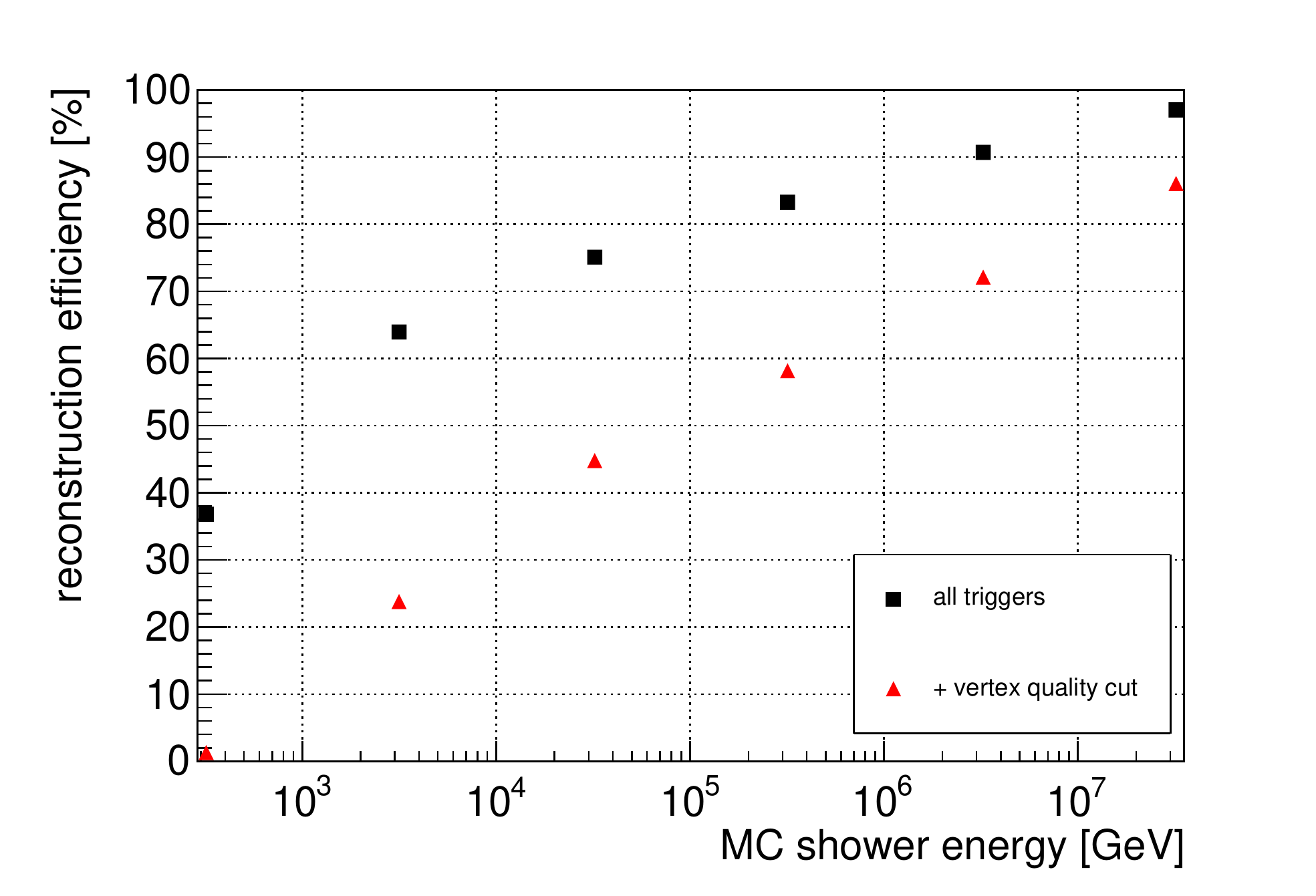}
 \caption{Left: The neutrino effective area after applying the
   vertex-quality cut to triggered events, and integrated over all
   directions, as a function of simulated neutrino energy for $\anue$
   (black full squares) and $\nue$ (red open squares) CC events, and
   $\overbar{\upnu}$ (black triangles) and for $\upnu$ (red open
   triangles) NC events.  Right: Reconstruction efficiency for all
   triggered shower-like events (black squares) and including the
   vertex-quality cut (red triangles) as a function of MC shower
   energy.}
\label{fig:reco-eff}
\end{figure*}
The effective area for the detection of $\nuane$ CC and all-flavour NC
events after applying the vertex-quality cut is depicted as a function
of the simulated neutrino energy in \myfref{fig:reco-eff} (left). The
peak in the effective area at roughly 6\,PeV for $\anue$ corresponds
to the Glashow resonance \cite{Glashow:1960zz}.  As shown in
\myfref{fig:reco-eff} (right), the fraction of successfully
reconstructed events among all triggered shower-like events increases
from $50\,\%$ to $90\,\%$ as a function of shower energy in the range
from $1\,\mathrm{TeV}$ to $3\,\mathrm{PeV}$.  Applying the
vertex-quality cut, roughly $10\,\%$ to $70\,\%$ of all triggered
shower-like events remain for the same energy range, while the
atmospheric muon background is reduced by 5 orders of magnitude.  The
remaining atmospheric muons are reconstructed with a mean zenith-angle
error of about $7^\circ$. Further details can be found in
Ref.~\cite{folger_phd}.

\section{Data selection}
\label{sec:dataselection}

The reconstruction algorithm described in the previous section was
applied to data collected from February 2007 to December 2012. This
includes the construction and commissioning phase of the detector and
therefore several detector configurations, each comprising a different
number of active lines included in the data taking.  All of these
configurations have been reproduced by the detailed run-based
simulation procedure described in \mysref{sec:simulation}.  The data
analysis was designed blindly,
i.e.\ the neutrino selection criteria have been developed using the
simulations only.  A fraction of 10\,\% of the data runs (test data),
sampled from the full data collection time range, 
was compared to simulations to validate the selection criteria.
These test data were excluded from the neutrino
search described in \mysref{sec:analysismethod}.  Simulation studies,
as well as a comparison to the test data, did not reveal any
significant influence of the time-variable optical background rates on
the performance of the shower reconstruction strategy presented in
\mysref{sec:reconstruction}.  This is to be expected, as the typical
optical background rates in the ANTARES detector are of the order of
50\,kHz to 80\,kHz per PMT, while even for extreme and rare conditions
of several hundred kHz, the probability of any given PMT having a
background hit in $\pm80\,\mathrm{ns}$ is of the order of a few
percent.

Active PMTs have been observed to occasionally produce a flash of
light inside OMs, and photons from this flash are detected by other
PMTs in the vicinity. This phenomenon is rare, with only a few
occurrences over the whole data-taking period.  Runs that have been
identified to contain at least one flashing PMT were excluded
from the analysis.  In order to further suppress this background,
events were vetoed if the shower position is reconstructed closer than
$15\,\text{m}$ to any of the OMs.  This cut ({\it discharge cut}),
which reduces the sensitive volume within the instrumented detector by
about 30\,\%, was chosen conservatively after a dedicated
analysis of events with flashing PMTs. Note that this cut is not
included in the effective area shown in \myfref{fig:reco-eff} (left).

Removing the 10\,\% test data, a total effective data-ac\-qui\-si\-tion
livetime of 1247 days is included in the analysis.

\section{Analysis method and systematic uncertainties}
\label{sec:analysismethod}
The presented analysis used 6 years of ANTARES data to search for an
excess over the atmospheric background of upward-going astrophysical
neutrinos inducing high-energy showering events.

The method is complementary to the first searches for a diffuse
neutrino flux performed with ANTARES
\cite{Aguilar:2010ab,Schnabel:2015ena}, which selected only the
track-like event signatures of upward-going muons induced by $\nuanmu$
CC interactions.  Even though NC interactions of atmospheric $\nuanmu$
contribute to the background for the presented search, the small value
of the ratio of atmospheric $\nuane$ to $\nuanmu$ fluxes at TeV
energies \cite{honda_atmflux} reduces the overall background compared
to the earlier analyses.


We treat the search for astrophysical neutrinos as a simple counting
experiment, and derive confidence intervals using the unified approach
of Feldman and Cousins \cite{feldmancousins}. We optimise the
selection criteria for the best upper limit, also known as model
rejection factor (MRF) optimisation \cite{modelrejection}.

Requiring successfully reconstructed shower-like events with hits on
at least 3 lines, which survive the vertex-quality (cf.\
\mysref{sec:reconstruction}) and the discharge cut (cf.\
\mysref{sec:dataselection}), reduces the atmospheric muon background
in the simulated event sample down to about $1000$ events, and about
$100$ ($10$) atmospheric (cosmic) neutrinos remain in the sample.

Selecting only upward-going shower events by cutting on their
reconstructed zenith angle, $\upTheta_{\text{rec}}=0^{\circ}$ defines
vertically down-going while $\upTheta_{\text{rec}}=180^{\circ}$ is
straight up-going, reduces this contamination further by a factor of
about 50.  Cutting on the reconstructed shower energy,
$\text{E}_{\text{rec}}$, in principle allows for the discrimination of
astrophysical and atmospheric neutrino contributions to the flux,
since the energy spectrum of astrophysical neutrinos is expected to be
harder than that of atmospheric neutrinos.

The MRF is minimised for a neutrino energy spectrum with spectral
index $\upGamma=2.0$ by varying $\text{E}_{\text{rec}}$ and
$\upTheta_{\text{rec}}$, and the optimum is obtained for
$\text{E}_{\text{rec}}\ge 10\,\text{TeV}$ and
$\upTheta_{\text{rec}} \ge 94^\circ$. 
It is found that this cut
combination vetoes the last simulated atmospheric muon events, and
that it is largely independent of the exact spectral shape of the
neutrino signal, in particular for softer spectral indices.
With
these cuts applied, the simulations yield an expectation of $1.3$ to
$2.9$ signal events ($\nuane+\nuanmu$) from a diffuse astrophysical
flux with the spectral index and normalisation as reported by IceCube
in Ref.~\cite{Aartsen:2016xlq} and Ref.~\cite{icecube:HESE4},
respectively.

In the following, all reported event contributions are given for the
cut level after the MRF optimisation.  From the simulated atmospheric
background, $2.3$ events are expected using the Bartol atmospheric
neutrino flux model~\cite{bartol_atmflux} and $0.3$ events from the
prompt atmospheric neutrino component. The latter assumes a flux
corresponding to the upper limit determined in
Ref.~\cite{Aartsen:2016xlq}, i.e. 50\,\% of the flux predicted in
Ref.~\cite{enberg_prompt}.  As no simulated atmospheric muon remains,
the residual contamination of atmospheric muons reconstructed as
upward-going showers is estimated by an extrapolation scheme.  The
efficiency of the vertex-quality cut applied on the sample of events
that survive the energy and zenith-angle cut was evaluated as a
function of the vertex-quality cut and was extrapolated to the strict
cut value used for the final event selection.  The validity of this
extrapolation scheme has been confirmed with looser cuts on the zenith
angle which allowed to compare with the number of muons remaining in
the sample.  This yields an estimate on the remaining atmospheric muon
contribution of $1.8$ events after the final cuts.

The contribution from astrophysical $\nuantau$ was estimated assuming
flavour equipartition at Earth for the astrophysical neutrino signal.
In the NC channel, $\nuantau$ interactions are assumed to create
showers identical to those of $\nuanmu$ and $\nuane$ interactions.
The contribution of $\nuantau$ CC interactions was estimated from the
$\nuane$ channel, taking into account that a fraction of $82.6\,\%$ of
all created $\uptau^{\pm}$ leptons will give rise to particle showers
through their decay.  This procedure estimates a total astrophysical
$\nuantau$ contribution of $0.5$ to $1.2$ events for the fluxes in
Ref.~\cite{Aartsen:2016xlq} and Ref.~\cite{icecube:HESE4}, with an
uncertainty of about 30\,\%, taking into account that the
$\uptau^{\pm}$ track length before decay exceeds the median vertex
resolution of the presented reconstruction for $\uptau^{\pm}$ energies
above roughly $100\,\mathrm{TeV}$, and can thus affect the shower fit.
The contribution of prompt atmospheric $\nuantau$ is negligible
\cite{enberg_prompt}.

For $\nuanmu$ CC events with $\mathrm{E}_{\upnu} > 100\,\text{TeV}$,
photon emission from the hadronic vertex shower has not been
simulated, cf. \mysref{sec:simulation}.  A dedicated analysis of the
reconstructed energy spectrum of such events for energies above and
below $100\,\text{TeV}$ was used to quantify their additional
contribution to the sample of reconstructed shower events.  This
estimate yields a small additional contribution of at most $0.3$
($0.2$) events from the astrophysical (atmospheric) $\nuanmu$ flux.

The systematic uncertainty on the normalisation of the conventional
atmospheric neutrino flux was assumed to be $\pm 30\,\%$
\cite{honda_atmflux,Aartsen:2013eka}.  The same was assumed as the
relative uncertainty on the number of atmospheric muons.  The
parametrisation in Ref.~\cite{enberg_prompt} was employed for the
prompt atmospheric neutrino flux which yields on average an
uncertainty of ${}^{+25}_{-40}\,\%$.

The influence of the uncertainty on the light absorption length and
the scattering length of seawater, and on the average PMT efficiency
has been determined by varying the nominal parameter values in the
detector simulation independently by
$\pm10$\,\%~\cite{Aguilar:2010kg}.  The resulting individual
uncertainties for the event detection efficiencies were added in
quadrature.  The number of simulated events surviving all cuts
relevant for the diffuse neutrino flux search, the assumed
uncertainties on the respective fluxes and the detection uncertainties
for the different fluxes are summarised in \mytref{tab:systematics}.
Neutrino events generated according to a hard astrophysical spectrum
are on average more energetic and hence induce a larger number of
signal hits in the detector compared to atmospheric neutrino events,
and their respective detection uncertainties are therefore smaller.

The uncertainty induced by the missing photon scattering in the
simulation of shower events has been investigated by a dedicated
simulation including photon scattering processes.  It was found that
on average $30\,\%$ less shower events with simulated photon
scattering survive the vertex-quality cut, which is taken into account
as a systematic uncertainty on the number of shower events in the
following.

\renewcommand{\arraystretch}{1.4}
\begin{table*}[]
\centering
    \begin{tabular}{ | l | c || c | c | p{3cm} |}
      \hline
      \multicolumn{2}{|c||}{events selected by final cuts} 
      & \multicolumn{2}{c|}{syst. uncertainties} \\ 
      \hline
      type 
      & number 
      & flux 
      & detection\\
      \hline
      conventional atmospheric $\nuane+\nuanmu$ 
      & 2.3
      & $\pm$30\,\%  
      & ${}^{+17}_{-23}\,\%$ \\
      
      + hadr. vertex corr. for $\text{E}_{\numu}>100$\,TeV
      & $\leq$0.2
      & 
      &  \\
      \hline
      prompt atmospheric $\nuan$
      & 0.3 
      & ${}^{+25}_{-40}\,\%$ 
      & --\\
      \hline 
      atmospheric $\upmu$
      & 1.8    
      & $\pm$30\,\%  
      & ${}^{+21}_{-22}\,\%$ \\
      \hline
      astrophysical $\nuane+\nuanmu$
      & 1.3 -- 2.9
      & --
      & ${}^{+14}_{-10}\,\%$ \\

      + hadr. vertex corr. for $\text{E}_{\numu}>100$\,TeV
      & $\leq$0.3
      & 
      &  \\
      \hline
      astrophysical $\nuantau$
      & 0.5 -- 1.2
      & --
      & $\pm$30\,\% \\
      \hline 
    \end{tabular}
    \caption{Event number expectations corresponding to 1247 days of
      data taking for the diffuse neutrino flux search
      derived from simulations for
      signal and background events. The range for the
      astrophysical event numbers corresponds to the fluxes as
      reported in Ref.~\cite{Aartsen:2016xlq}
      and Ref.~\cite{icecube:HESE4}, respectively.
      Event numbers for a given neutrino flavour denote
      the sum of neutrinos and their respective antineutrinos.
      Additionally, the assumed systematic uncertainties on the
      fluxes, 
      and uncertainties on the detection efficiency, as inferred from
      detector simulations after the vertex-quality cut only
      (cf. \mysref{sec:reconstruction}), are shown. 
    }
    \label{tab:systematics}
\end{table*}

\section{Results}
\label{sec:results}

Summing up the discussed atmospheric background contributions and
correction estimates (cf. \mytref{tab:systematics}),
$\text{n}_{\text{b}}=4.6^{+2.8}_{-3.0}$ background events are
expected.  For the full dataset of 1247 days, this analysis yields a
sensitivity to an astrophysical neutrino flux of:
\begin{equation*}
\label{finalsensitivity}
\text{E}^2 \cdot \bar \upPhi^{\text{90\%}} = 2.2^{+0.9}_{-0.7} \cdot 10^{-8} \, \diffunit
\end{equation*}
per flavour, assuming an unbroken $\text{E}^{-2}$ power law spectrum
and flavour equipartition at Earth.
\begin{figure}[ht!]
\centering
\includegraphics[width=0.99\linewidth]{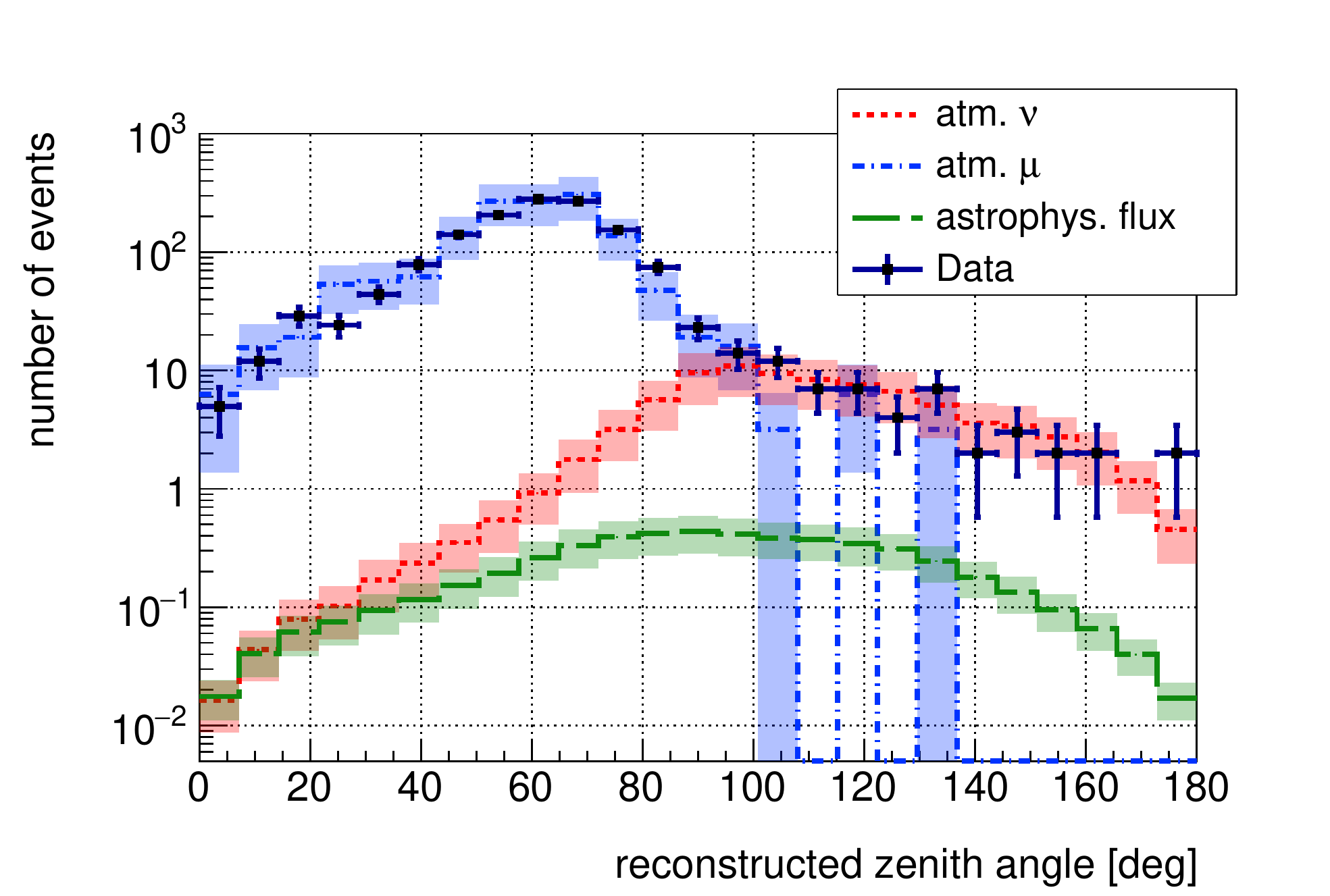}
\caption[Comparison of MC simulations and ANTARES data]{
  Reconstructed zenith-angle distribution for 1247 days of data taking,
  with events selected as described in \mysref{sec:reconstruction} and
  \mysref{sec:dataselection}. Data points and their statistical errors
  are depicted with black markers and compared to simulated
  distributions of atmospheric muons (blue), atmospheric neutrinos
  (red) and the astrophysical flux reported in Ref.~\cite{Aartsen:2016xlq}
  (green). The coloured bands indicate the uncertainties on the
  simulated and measured flux normalisations.}
\label{fig:zenith}
\end{figure}

\myfref{fig:zenith} shows the reconstructed zenith-angle distribution.
The cuts discussed in \mysref{sec:reconstruction} and
\mysref{sec:dataselection} were applied. The measured distribution
compares well to the MC expectations from the atmospheric muon and
neutrino backgrounds. The zenith-angle distribution of the atmospheric
neutrino background is asymmetric with respect to the horizon, which
results from the convolution of the assumed atmospheric neutrino flux
model~\cite{bartol_atmflux} and the detector acceptance.

Applying a cut on the reconstructed zenith angle
$\upTheta_{\text{rec}} \ge 94^\circ$, as derived in the MRF optimization
procedure, 60 upward-going events remain, while 35 have a
reconstructed shower position inside the instrumented volume.  As
expected from simulations, the remaining are reconstructed at a
maximum distance of $84\,\mathrm{m}$ to the surface of the volume
enclosed by the detector lines.
\begin{figure}[ht!]
\centering
\includegraphics[width=0.99\linewidth]{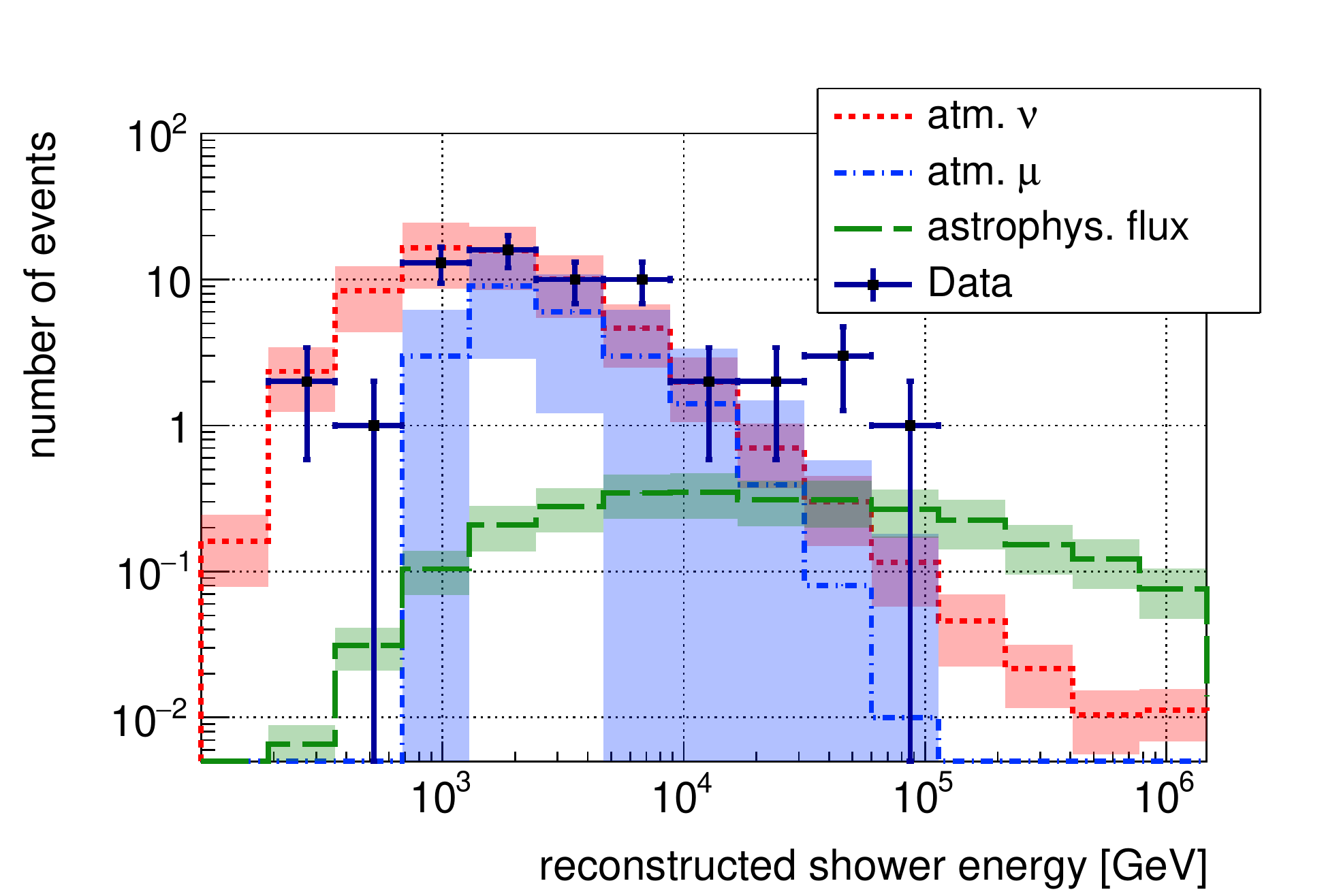}
\caption[Comparison of MC simulations and ANTARES data]{ Distribution
  of the reconstructed shower energy for 1247 days of data taking,
  selected as described in \mysref{sec:reconstruction} and with a cut
  on the reconstructed zenith angle applied at
  $\upTheta_{\text{rec}} \ge 94^\circ$ (black markers, statistical
  errors only).  Simulated contributions from atmospheric muons
  (blue), atmospheric neutrinos (red) and an astrophysical flux
  ~\cite{Aartsen:2016xlq} (green) have been overlaid for
  comparison. Coloured bands indicate the uncertainties on the
  simulated and measured flux normalisations. The atmospheric muon
  contribution beyond 10\,TeV has been extrapolated as described in
  \mysref{sec:analysismethod}.}
\label{fig:energy}
\end{figure}
\myfref{fig:energy} depicts the reconstructed energy spectrum of these
60 events, again compared to expectations derived from simulations.
Applying the additional and final cut on the reconstructed shower
energy $\text{E}_{\text{rec}}\ge 10\,\mathrm{TeV}$ results in 8
remaining events. All of these events have their shower vertex
position reconstructed outside of the instrumented volume.  Each of
these 8 events has been investigated individually by a dedicated
event-based MC simulation.  One event was identified to have surpassed
the $\geq 3$ line veto criterion (cf. \mysref{sec:reconstruction}) due
to 2 additional, isolated random hits on 2 different lines which
coincidentally matched to the shower hypothesis.  This is a scenario
which is in principle covered by the run-based simulation concept that
accounts for the OM-individual background rates at the time of the
data taking.  The remaining 7 events could be verified to have a
reconstruction error comparing well to the resolutions discussed in
\mysref{sec:reconstruction}.

Using Poisson statistics, the observation of 8 events with an
expectation of $4.6$ corresponds to an excess with a significance of
$1.6\,\upsigma$.  This result agrees with the assumption of a purely
atmospheric origin of the observed events, but it is also compatible
with the expectations from the diffuse astrophysical neutrino fluxes
as reported by the IceCube collaboration.

Following the Feldman-Cousins approach~\cite{feldmancousins} a
$90\,\%$ C.L. upper limit on the number of signal events of
$\upmu_{\text{90\%}} = 9.1$ is evaluated from the 8 measured and
$\text{n}_{\text{b}}=4.6^{+2.8}_{-3.0}$ expected background events.
Systematic uncertainties (including that arising from the missing
photon scattering in our simulation, cf. \mysref{sec:analysismethod})
have been taken into account following the method detailed in
Refs.~\cite{conrad_confidenceintervals,conrad_pole}.

The relative uncertainties on the signal and background efficiencies,
calculated as the average of their systematic error intervals, are
evaluated to $29\,\%$ for the astrophysical signal and $42\,\%$ for
the atmospheric background.  This increases the 90\,\% C.L. upper
limit of the confidence interval to $11.4$ events.  For the unblinded
data set of 1247 days, the upper limit on the diffuse astrophysical
neutrino flux per neutrino flavour is then evaluated to:
\begin{equation*} 
\text{E}^2 \cdot \upPhi^{90\%} = 4.9 \cdot 10^{-8} \, \diffunit.
\end{equation*} 
The limit is valid under the assumption of flavour equipartition at
Earth and for an unbroken $\text{E}^{-2}$ spectrum in the energy range
from $23\,\mathrm{TeV}$ to $7.8\,\mathrm{PeV}$. This range was
obtained from the simulated neutrino energy spectrum of all
astrophysical shower-like events by determining its central 90\,\%
interval.

\section{Summary and Conclusion}
\label{sec:conclusion}

A novel event reconstruction algorithm has been presented, which
allowed for the first time to select and reconstruct particle-induced
shower events in data taken with the ANTARES neutrino telescope.  The algorithm
achieves a median angular resolution of $6^\circ$ for shower energies
below 100\,TeV. The median value of the true over
reconstructed shower energy ratio, $\mathrm{E_{MC}}/\mathrm{E_{rec}}$,
is 1.5 -- 2 for shower energies up to 1\,PeV, while the 90\% quantile
increases from 3 to 20 for energies between 500\,GeV and 1\,PeV.  The
fraction of successfully reconstructed events among all triggered
shower-like events increases from $50\,\%$ to $90\,\%$ as a function
of the shower energy in the range from 1\,TeV to
3\,PeV.

Using 1247 days of ANTARES data, a $90\,\%$ C.L. upper limit on a
diffuse astrophysical neutrino flux per flavour was evaluated to:
\begin{equation*}
\label{summaryupperlimit}
\text{E}^2 \cdot \upPhi^{90\%} = 4.9 \cdot 10^{-8} \, \diffunit .
\end{equation*}
The limit is valid in the energy range from $23\,\mathrm{TeV}$ to
$7.8\,\mathrm{PeV}$, assuming an unbroken $\text{E}^{-2}$ neutrino
spectrum and flavour equipartition at Earth.  It has been calculated
using the Feldman-Cousins approach~\cite{feldmancousins}.  Systematic
errors have been taken into account following
Refs.~\cite{conrad_confidenceintervals,conrad_pole}.

\begin{figure}[t]
\centering
\resizebox{0.99\textwidth}{!}{%
  \includegraphics{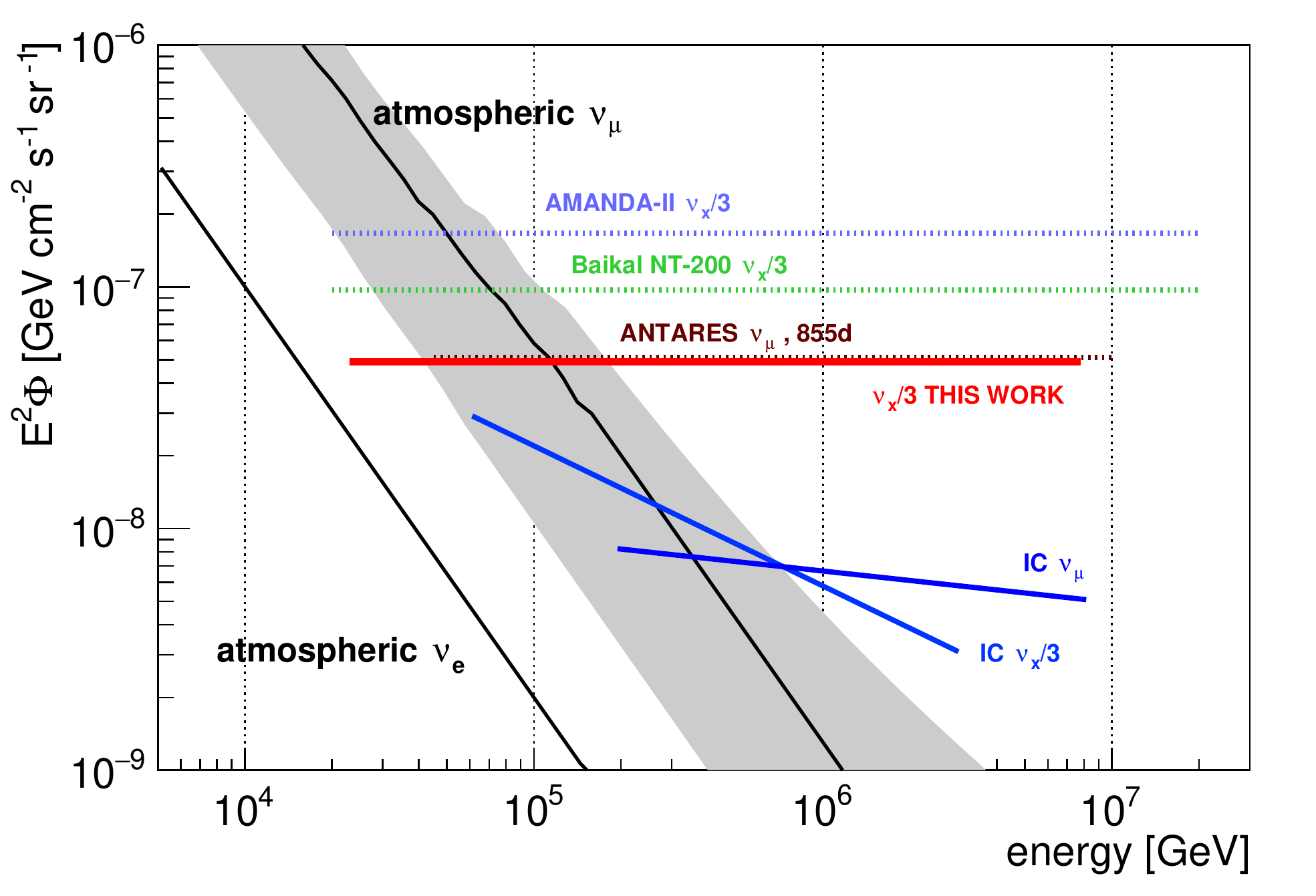}
}
\caption[The result of this work in the context of previous analyses.]
{The 90\% C.L. upper limit on the diffuse all-flavour astrophysical
  neutrino flux obtained in this work (solid red line) in comparison
  to previously set upper limits (dotted lines, AMANDA-II
  ~\cite{amanda_diffuse}, Baikal NT-200~\cite{baikal_icrc09diffuse},
  and ANTARES $\numu$~\cite{Schnabel:2015ena}) and 2 different
  measurements of a diffuse astrophysical neutrino flux reported by
  IceCube (solid blue lines, IC $\upnu_{\text{x}}/3$
  ~\cite{icecube:HESE4}, and IC $\numu$~\cite{Aartsen:2016xlq}).}
\label{fig:limits}
\end{figure} 

\myfref{fig:limits} illustrates the obtained upper limit in comparison
with previously set limits by the AMANDA \cite{amanda_diffuse} and
Baikal~\cite{baikal_icrc09diffuse} experiments.  The upper limit
obtained in this work almost coincides with those obtained previously
with ANTARES, using only upward-going muons recorded in
855 \cite{Schnabel:2015ena} and 334 \cite{Aguilar:2010ab} days,
although the sensitivity of the present dataset is about a factor of
two and three better, respectively.  Also shown are the two most
recent IceCube measurements of an astrophysical flux that have been
obtained either with analyses selecting contained
events~\cite{icecube:HESE4} or using through-going muon tracks
originating from the Northern sky~\cite{Aartsen:2016xlq}.  All flux
limits and measurements are given per flavour and represent the sum of
neutrino and antineutrino fluxes.  For comparison, the conventional
atmospheric $\numu$ flux (black solid line with the gray shaded area
showing systematic uncertainty) according to the Bartol neutrino flux
model~\cite{bartol_atmflux} and the measured atmospheric $\nue$
flux~\cite{PhysRevD.91.122004} is indicated.

The reported measurement of $8$ events is statistically in agreement
with the expected background of $4.6^{+2.8}_{-3.0}$ events from
atmospheric muons and neutrinos.  Assuming an astrophysical flux as
reported in Ref.~\cite{Aartsen:2016xlq} (\cite{icecube:HESE4}),
additional $2.1$ ($4.4$) signal events are expected, which reduces to
$1.7$ ($4.2$) events assuming a cut-off at 3\,PeV.  In all cases, the
addition of an astrophysical neutrino signal is compatible with our
measurement.

Though not yet sufficiently sensitive, the presented first shower
analysis using the initial 6 years of data taken with the ANTARES
neutrino telescope demonstrates the potential of ANTARES to
independently confirm and complement the measurement of a high-energy
astrophysical neutrino flux, as performed by IceCube. In order to meet
this important goal, several improvements of the analysis have been
identified and are under way. Building on the gained experience, a
second shower reconstruction strategy is developed. It improves on the
angular resolution and increases the shower event selection
efficiency, while continuing to provide the necessary strong
suppression of the atmospheric muon background. Using the track
reconstruction already employed in our previous searches for a diffuse
flux with muon neutrinos~\cite{Aguilar:2010ab,Schnabel:2015ena}, an
analysis combining both track-like and shower-like events is in
progress.  With the addition of the remaining ANTARES data until the
scheduled end of its operation time in 2017, this combined search is
expected to reach a sensitivity at the level of the flux discovered by
IceCube \cite{Aartsen:2014muf}.

\section{Acknowledgements}
\label{sec:acknowledgment}
The authors acknowledge the financial support of the funding agencies:
Centre National de la Recherche Scientifique (CNRS), Commissariat \`a
l'\'ener\-gie atomique et aux \'energies alternatives (CEA),
Commission Europ\'eenne (FEDER fund and Marie Curie Program),
Institut Universitaire de France (IUF), IdEx program and UnivEarthS
Labex program at Sorbonne Paris Cit\'e (ANR-10-LABX-0023 and
ANR-11-IDEX-0005-02), Labex OCEVU (ANR-11-LABX-0060) and the
A*MIDEX project (ANR-11-IDEX-0001-02),
R\'egion \^Ile-de-France (DIM-ACAV), R\'egion
Alsace (contrat CPER), R\'egion Provence-Alpes-C\^ote d'Azur,
D\'e\-par\-tement du Var and Ville de La
Seyne-sur-Mer, France;
Bundesministerium f\"ur Bildung und Forschung
(BMBF), Germany; 
Istituto Nazionale di Fisica Nucleare (INFN), Italy;
Stichting voor Fundamenteel Onderzoek der Materie (FOM), Nederlandse
organisatie voor Wetenschappelijk Onderzoek (NWO), the Netherlands;
Council of the President of the Russian Federation for young
scientists and leading scientific schools supporting grants, Russia;
National Authority for Scientific Research (ANCS), Romania;
Mi\-nis\-te\-rio de Econom\'{\i}a y Competitividad (MINECO): Plan
Estatal de Investigaci\'{o}n (refs. FPA2015-65150-C3-1-P, -2-P and
-3-P, (MINECO / FEDER)), Severo Ochoa Centre of Excellence and
MultiDark Consolider (MINECO), and Prometeo and Grisol\'{i}a programs
(Generalitat Valenciana), Spain;
Ministry of Higher Education, Scientific Research and Professional Training, Morocco.
We also acknowledge the technical support of Ifremer, AIM and Foselev Marine
for the sea operation and the CC-IN2P3 for the computing facilities.

\providecommand{\etal}{et al.\xspace}
\providecommand{\coll}{Coll.\xspace}

\end{document}